\documentclass[preprint,aps]{revtex4}
\usepackage{graphicx,amsmath}
\begin{document}
\newcommand{\etal}{{\it{et al.}}~}
\newcommand{\bqe}{\begin{equation}}
\newcommand{\eqe}{\end{equation}}
\newcommand{\bqs}{\begin{equation*}}
\newcommand{\eqs}{\end{equation*}}
\newcommand{\bqa}{\begin{eqnarray}}
\newcommand{\barint}{-\hspace{-.45cm}\int}
\newcommand{\barinta}{-\hspace{-.37cm}\int}
\newcommand{\eqa}{\end{eqnarray}}
\newcommand{\bqas}{\begin{eqnarray*}}
\newcommand{\eqas}{\end{eqnarray*}}
\newcommand{\nn}{\noindent}
\newcommand{\no}{\nonumber}
\newcommand{\ssi}{\scriptsize}

\title[Optimizing snake locomotion in the plane. II. Large transverse friction]{
Optimizing snake locomotion in the plane. II. Large transverse friction}

\author{Silas Alben}

\affiliation{Department of Mathematics, University of Michigan,
Ann Arbor, MI, USA}
\label{firstpage}

\begin{abstract}
We determine analytically the form of optimal snake locomotion when
the coefficient of transverse friction is large, the typical
regime for biological and robotic snakes. We
find that the optimal snake motion is a retrograde
traveling wave, with a wave amplitude that decays as
the -1/4 power of the coefficient of transverse friction. This
result agrees well with our numerical computations.
\end{abstract}

\maketitle
\section{Introduction}

In \cite{AlbenSnake2013I} we considered the problem of optimizing
snake locomotion in the plane numerically. The mechanics
of snake locomotion has been described previously by biologists
and engineers, and modeled by applied mathematicians
\citep{gray1950kinetics,GuMa2008a,HuNiScSh2009a,HaCh2010b,MaHu2012a,HuSh2012a}.
As with other terrestrial locomoting animals \citep{dickinson2000animals},
a terrestrial snake pushes against the ground with its
body, and obtains a reaction force from the ground which
propels it forward. However, a wide range of possible kinematics
can be employed, and determining which are most effective
(i.e. efficient) in different environments, and why, has been
a major theme of locomotion studies \citep{BeKoSt2003a,AvGaKe2004a,TaHo2007a,fu2007theory,spagnolie2010optimal,crowdy2011two,
lighthill1975mathematica,childress1981mechanics,sparenberg1994hydrodynamic,alben2009passive,michelin2009resonance,peng2012bb}.

A first approximation to snake locomotion is to consider
motions confined to two dimensions
\cite{GuMa2008a,HuNiScSh2009a,HuSh2012a}.
Here the reaction force in the plane of snake motion is
due to friction, and Coulomb friction provides a simple
model. Hu and Shelley \cite{HuNiScSh2009a,HuSh2012a}
emphasized the importance of frictional anisotropy in
snake locomotion. In particular, they found that
when the curvature of the snake backbone is
prescribed as a sinusoidal traveling wave, 
high speed and efficiency is obtained when
the coefficient of transverse friction is large
compared to the coefficient of forward friction.
Jing and Alben \cite{JiAl2013} found that the same
holds for two- and three-link snakes. In biological
snakes, the ratio of transverse to forward friction
is thought to be greater than one \cite{HuNiScSh2009a,HuSh2012a},
although how much greater is not clear. In
\cite{HuNiScSh2009a,HuSh2012a}, a ratio of only 1.7 was
measured for anaesthetized snakes. These works
noted that active snakes use their scales to increase
frictional anisotropy, so the ratio in locomoting
snakes is potentially much larger, and indeed,
a ratio of 10 was found to give better agreement
between the Coulomb friction model and
a biological snake \cite{HuSh2012a}.
Wheeled snake
robots have been employed very successfully
for locomotion \cite{hirosebiologically,hopkins2009survey},
and for most wheels the frictional
coefficient ratio (with forward friction defined
by the rolling resistance coefficient of the wheel)
is much greater than 10 \cite{persson2000sliding}.

In this paper, we develop an analytical solution
for the optimal locomotion of a snake in the plane,
when the ratio of transverse to
forward friction is large. We make few assumptions on the snake
kinematics at the outset, but use numerical computations
from \cite{AlbenSnake2013I} to provide guiding
intuition. We find first that the optimal form of
the prescribed backbone curvature is
a traveling wave. We then find that the
root-mean-square (RMS) amplitude of the
 curvature should decay as
the $-1/4$ power of the ratio of transverse
to forward friction coefficients. Surprisingly,
any periodic traveling wave motion can achieve
the optimal efficiency, subject to the
aforementioned restriction on its
RMS amplitude, and to the requirement that
its wavelength, normalized by the snake
length, tend to zero. In the limit of
large transverse friction, the power
required to move a
snake optimally is simply that needed to
tow a straight snake forward.

\section{Model \label{sec:Model}}

\begin{figure} [h]
           \begin{center}
           \begin{tabular}{c}
               \includegraphics[width=5in]{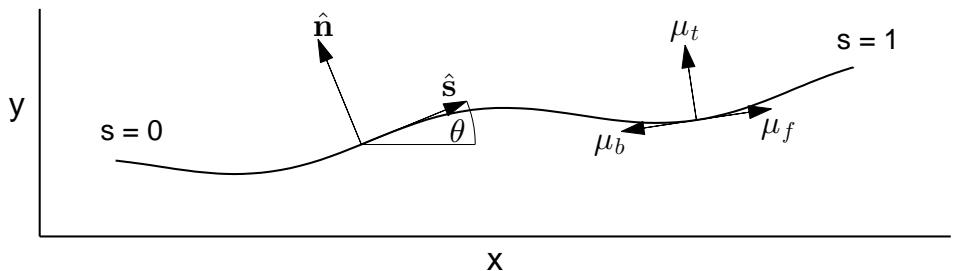} \\
           \vspace{-.25in}
           \end{tabular}
          \caption{\footnotesize Schematic diagram of snake position,
parametrized by arc length $s$ (nondimensionalized by snake length),
 at an instant in time.  The tangent angle
and unit vectors
tangent and normal to the curve at a point are labeled. Vectors representing
forward, backward and transverse velocities are shown with
the corresponding friction coefficients $\mu_f$, $\mu_b$, and $\mu_t$.
 \label{fig:SnakeSchematic}}
           \end{center}
         \vspace{-.10in}
        \end{figure}

We use the same frictional
snake model as \citep{HuNiScSh2009a,HuSh2012a,JiAl2013,AlbenSnake2013I}, so we only summarize it
here. The snake's
position is given by $\mathbf{X}(s,t) = (x(s,t), y(s,t))$, a planar curve which is
parametrized by arc length $s$ and varies with time $t$. A schematic
diagram is shown in figure \ref{fig:SnakeSchematic}.

The unit vectors
tangent and normal to the curve are $\hat{\mathbf{s}}$ and $\hat{\mathbf{n}}$
respectively. The tangent angle and curvature are denoted $\theta(s,t)$ and $\kappa(s,t)$,
and satisfy $\partial_s x = \cos\theta$, $\partial_s y = \sin\theta$, and
$\kappa = \partial_s \theta$. We consider the problem of prescribing the curvature of
the snake as a function of time, $\kappa(s,t)$, in order to obtain efficient locomotion.
When $\kappa(s,t)$ is prescribed, the tangent angle and position are obtained by integration:
\begin{align}
\theta(s,t) &= \theta_0(t) + \int_0^s \kappa(s',t) ds', \label{theta0} \\
x(s,t) & = x_0(t) + \int_0^s \cos \theta(s',t) ds', \label{x0}\\
y(s,t) &= y_0(t) + \int_0^s \sin \theta(s',t) ds'. \label{y0}
\end{align}
\nn The trailing-edge position $(x_0, y_0)$ and tangent angle $\theta_0$
are determined by the force and torque
balance for the snake, i.e. Newton's second law:
\begin{align}
\int_0^L \rho \partial_{tt} x ds &= \int_0^L f_x ds, \label{fx0} \\
\int_0^L \rho \partial_{tt} y ds &= \int_0^L f_y ds, \label{fy0} \\
\int_0^L  \rho \mathbf{X}^\perp \cdot \partial_{tt} \mathbf{X} ds
&= \int_0^L \mathbf{X}^\perp \cdot \mathbf{f} ds. \label{torque0}
\end{align}
\nn Here $\rho$ is the snake's mass per unit length and $L$ is
the snake length. The snake is locally inextensible,
and $\rho$ and $L$ are constant in time.
$\mathbf{f}$ is the force per unit length on the snake
due to Coulomb friction with the ground \cite{HuNiScSh2009a}:
\bqe
\mathbf{f}(s,t) = -\rho g \mu_t
\left( \widehat{\partial_t{\mathbf{X}}}\cdot \hat{\mathbf{n}} \right)\hat{\mathbf{n}}
-\rho g \left(\mu_f H(\widehat{\partial_t{\mathbf{X}}}\cdot \hat{\mathbf{s}})
+ \mu_b (1-H(\widehat{\partial_t{\mathbf{X}}}\cdot \hat{\mathbf{s}}))\right)
\left( \widehat{\partial_t{\mathbf{X}}}\cdot \hat{\mathbf{s}} \right)\hat{\mathbf{s}}. \label{friction}
\eqe
\nn Here $H$ is the Heaviside function and the hats denote
normalized vectors. When $\|\partial_t{\mathbf{X}}\| = \mathbf{0}$
we define $\widehat{\partial_t{\mathbf{X}}}$ to be $\mathbf{0}$.
According to (\ref{friction}) the snake experiences
friction with different coefficients for motions in different directions.
The frictional coefficients are $\mu_f$, $\mu_b$, and $\mu_t$ for motions
in the forward ($\hat{\mathbf{s}}$), backward ($-\hat{\mathbf{s}}$),
and transverse (i.e. normal) directions ($\pm\hat{\mathbf{n}}$), respectively. In
general the snake velocity at a given point has both tangential and
normal components, and the frictional force density has
components acting in each direction. A similar decomposition of force
into directional components
occurs for viscous fluid forces on slender bodies \citep{cox1970motion}.

We assume that the snake curvature $\kappa(s,t)$ is a prescribed function of
$s$ and $t$ that is periodic in $t$ with period $T$. Many of the motions
commonly observed in real snakes are essentially periodic in time \citep{HuNiScSh2009a}.
We nondimensionalize equations (\ref{fx0})--(\ref{torque0}) by dividing
lengths by the snake length $L$, time by $T$, and mass by $\rho L$. Dividing
both sides by $\mu_f g$ we obtain:
\begin{align}
\frac{L}{\mu_f gT^2} \int_0^1 \partial_{tt} x ds &= \int_0^1 f_x ds, \label{fxa} \\
\frac{L}{\mu_f gT^2}\int_0^1 \partial_{tt} y ds &= \int_0^1 f_y ds, \label{fya} \\
\frac{L}{\mu_f gT^2}\int_0^1 \mathbf{X}^\perp \cdot \partial_{tt} \mathbf{X} ds
&= \int_0^1 \mathbf{X}^\perp \cdot \mathbf{f} ds. \label{torquea}
\end{align}
\nn In (\ref{fxa})--(\ref{torquea}) and from now on, all variables are
dimensionless. For most of the snake motions observed in nature, $L/\mu_f gT^2 \ll 1$
\citep{HuNiScSh2009a}, which means that the snake's inertia is negligible. By setting
this parameter to zero we simplify the problem
considerably while maintaining
 a good representation of real snakes. (\ref{fxa})--(\ref{torquea}) become:
\begin{align}
\mathbf{b} = (b_1,b_2,b_3)^\top = \mathbf{0} \quad ; \quad b_1 &\equiv \int_0^1 f_x ds, \label{fxb} \\
b_2 &\equiv \int_0^1 f_y ds, \label{fyb} \\
b_3 &\equiv \int_0^1 \mathbf{X}^\perp \cdot \mathbf{f} ds. \label{torqueb}
\end{align}
\nn In (\ref{fxb})--(\ref{torqueb}), the dimensionless force $\mathbf{f}$ is
\bqe
\mathbf{f}(s,t) = -\frac{\mu_t}{\mu_f}
\left( \widehat{\partial_t{\mathbf{X}}}\cdot \hat{\mathbf{n}} \right)\hat{\mathbf{n}}
- \left( H(\widehat{\partial_t{\mathbf{X}}}\cdot \hat{\mathbf{s}})
+ \frac{\mu_b}{\mu_f} (1-H(\widehat{\partial_t{\mathbf{X}}}\cdot \hat{\mathbf{s}}))\right)
\left( \widehat{\partial_t{\mathbf{X}}}\cdot \hat{\mathbf{s}} \right)\hat{\mathbf{s}} \label{friction1}
\eqe
\nn The equations (\ref{fxb})--(\ref{torqueb}) thus involve only two parameters,
which are ratios of the friction coefficients. From now on, for simplicity, we
refer to $\mu_t/\mu_f$ as $\mu_t$ and $\mu_b/\mu_f$ as $\mu_b$. Without
loss of generality, we assume $\mu_b \geq 1$. This amounts to defining
the backward direction as that with the higher of the tangential frictional
coefficients, when they are unequal. $\mu_t$ may assume any nonnegative value.
The same model was used in \citep{HuNiScSh2009a,HuSh2012a,JiAl2013},
and was found to agree well with the motions of biological
snakes in \citep{HuNiScSh2009a}.

Given the curvature $\kappa(s,t)$, we solve the three nonlinear equations
(\ref{fxb})--(\ref{torqueb}) at each time $t$ for the three unknowns
$x_0(t)$, $y_0(t)$ and $\theta_0(t)$. Then we obtain the snake's
position as a function of time by (\ref{theta0})--(\ref{y0}).
The distance
traveled by the snake's center of mass over one period is
\bqe
d = \sqrt{\left(\int_0^1 x(s,1) - x(s,0) \,ds\right)^2 +
\left(\int_0^1 y(s,1) - y(s,0)\, ds\right)^2}. \label{dist}
\eqe
The work done by the snake against friction over one period is
\begin{align}
W &= \int_0^1 \int_0^1 \mathbf{f}(s,t) \cdot \partial_t\mathbf{X}(s,t)\, ds\, dt \label{W}
\end{align}
\nn 
We define the cost of locomotion as
\bqe
\eta = \frac{W}{d} \label{eta}
\eqe
\nn and our objective here is to find $\kappa(s,t)$ which minimizes $\eta$
as $\mu_t \to \infty$.

\section{Large-$\mu_t$ analysis \label{LargeMut}}

\begin{figure} [h]
           \begin{center}
           \begin{tabular}{c}
               \includegraphics[width=6in]{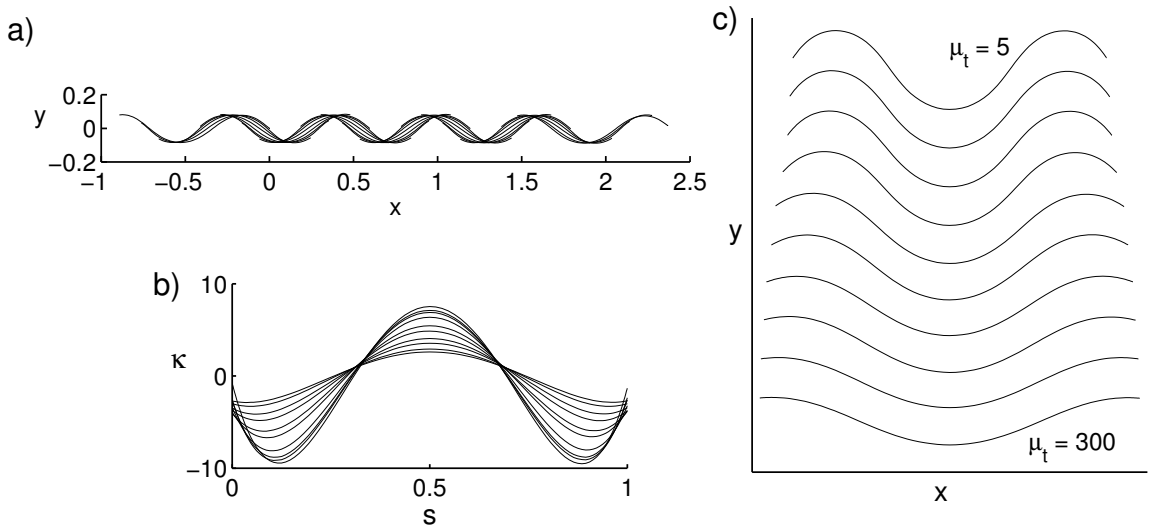} \\
           \vspace{-.25in}
           \end{tabular}
          \caption{\footnotesize Numerical traveling-wave optima
from \cite{AlbenSnake2013I} for ten different $\mu_t \gg 1$.
a) Snake trajectory over one period, from
one snake optimization run for $\mu_b = 1$ and $\mu_t = 30$. b) Curvature versus
arc length at the instant when a curvature maximum crosses the snake
midpoint, for ten $\mu_t$ values: 5, 6, 7, 10, 20, 30, 60, 100, 200, and 300.
c) Snake shapes corresponding to the curvatures in (b).
 \label{fig:HighMutPaperFig1}}
           \end{center}
         \vspace{-.10in}
        \end{figure}

We now analytically determine the optimal snake motion in the limit of large $\mu_t$.
Figure \ref{fig:HighMutPaperFig1} shows a few numerical results from \cite{AlbenSnake2013I},
which provide some intuition as we develop the analytical solution. In
\cite{AlbenSnake2013I} we show
that for $\mu_t \gtrsim 6$, the numerically-computed optimal motions
are always (retrograde) traveling waves, an example of
which is shown in panel a for $\mu_t = 30$. A sequence of snapshots of
the snake is shown over a period of motion. The snake moves from left
to right, and appears to
follow a sinusoidal path, very similar to what has been observed
biologically and studied numerically \cite{HuSh2012a}. We have computed similar traveling-wave
optima at a range of $\mu_t \gg 1$. In panel b we compare the curvatures from
these optimal motions at a particular instant when a curvature maximum occurs
at the snake midpoint, for ten different values of $\mu_t$: 5, 6, 7, 10, 20, 30, 60, 100, 200, and 300.
The profiles are
similar, but the amplitudes decay monotonically as $\mu_t$ increases. Panel c
shows the ten snake shapes corresponding to the curvatures in
panel b. Consistently, we see the decay of the traveling-wave amplitude as $\mu_t$
increases. With this picture from the numerics, we now proceed to derive
the analyical solution in the asymptotic limit of large $\mu_t$.

We begin with the position of the snake
in terms of its components:
\bqe
\mathbf{X}(s,t) = (x(s,t), y(s,t)). \label{Pos}
\eqe
\nn If $\mu_t$ is large, the snake moves more easily in the tangential direction than in
the transverse direction. Furthermore, the most efficient
curvature function will give motion mainly in the tangential direction, to avoid large work done
against friction. We assume that the mean direction of motion is aligned with the $x$-axis,
and that the deflections from the $x$-axis are small---that is, $|y|, |\partial_t y|,
|\partial_s y|$ and higher derivatives are $O(\mu_t^\alpha)$ for some negative $\alpha$.
This has been observed in the numerical solutions, and makes intuitive sense. Small deflections
allow the snake to move along a nearly straight path, which is more efficient than a more
curved path, which requires more {\it tangential} motion (and work done against
tangential friction) for a given {\it forward} motion, i.e. $d$.

We expand each of the terms in the force and torque balance equations in powers of
$|y|$ and retain only the terms which are dominant at large $\mu_t$. We expand the
cost of locomotion similarly, and find the dynamics which
minimize it, in terms of $y(s,t)$.

We first expand $x(s,t)$. We decompose $x(s,t)$ into its $s$-average $\bar{x}$ and a
zero-$s$-average remainder:
\bqe
x(s,t) = \overline{x(s,t)} + \barint^s \cos{\theta(s',t)}\,ds'
\eqe
\nn where $\barinta^s$ means the constant of integration is chosen
so that the integrated function has zero $s$-average. Hence
\bqe
\partial_t x(s,t) = \overline{\partial_t x(s,t)} + \barint^s -\partial_t \theta(s',t) \sin{\theta(s',t)}\,ds'. \label{Vel}
\eqe
\nn We denote the $s$-averaged horizontal velocity (the horizontal velocity of the snake center of mass) by
\bqe
U(t) \equiv \overline{\partial_t x(s,t)}.
\eqe
\nn We expand the integrand in (\ref{Vel}) in powers of $|y|$ and its derivatives:
\bqe
-\partial_t \theta(s,t) \sin{\theta(s,t)} = -\partial_{st} y(s,t) \partial_s y(s,t) + O(|\partial_s y|^4). \label{integrand}
\eqe
\nn The quartic remainder term in (\ref{integrand}), $O(|\partial_s y|^4)$, actually includes other
quartic terms involving time derivatives
of $\partial_s y$, but for brevity we use
the assumption (which will be correct for our solution) 
that $|y|$ and all its derivatives are of the same order
of smallness with respect to $\mu_t$. We define
\bqe
h(s,t) = \barint^s -\partial_{s't} y(s',t) \partial_{s'} y(s',t) ds' \label{h}
\eqe
\nn and with the small-deflection assumption, (\ref{Vel}) becomes
\bqe
\partial_t x(s,t) = U(t) + h(s,t) + O(|\partial_s y|^4), \label{Velxh}
\eqe
\nn and
\bqe
\partial_t\mathbf{X}(s,t) = (U(t) + h(s,t), \partial_t y(s,t))+ O(|\partial_s y|^4). \label{Velexp}
\eqe
\nn We will see that for a general class of shape dynamics near the optimum, 
$U(t)$ is $O$(1), i.e. of the same order as one snake length per actuation
period, even in the limit of large $\mu_t$. $U$ is therefore the dominant term in
(\ref{Velxh}) and (\ref{Velexp}). We use this assumption on $U$ to expand the velocity 2-norm:
\begin{align}
\| \partial_t\mathbf{X}(s,t) \| &= \sqrt{\partial_t x^2 + \partial_t y^2}
 = U\left(1 + \frac{h}{U} + \frac{1}{2}\frac{\partial_t y^2}{U^2}\right) + O(|y|^4). \label{Velnormexp}
\end{align}
\nn In writing $O(|y|^4)$ we have again lumped terms with four or more powers of $|y|$ and/or 
its $s$- and $t$-derivatives. Dividing
(\ref{Velexp}) by (\ref{Velnormexp}), we obtain the expansion for the normalized velocity:
\bqe
\widehat{\partial_t{\mathbf{X}}}(s,t) =
\left( {\begin{array}{c}
 1 - \dfrac{1}{2}\dfrac{\partial_t y^2}{U^2} + O(|y|^4) \vspace{0.3cm}\\
 \dfrac{\partial_t y}{U}\left(1 - \dfrac{h}{U} - \dfrac{1}{2}\dfrac{\partial_t y^2}{U^2}\right)  + O(|y|^5)  \\
 \end{array} } \right). \label{Velnormdexp}
\eqe
\nn We next expand the unit tangent and normal vectors:
\begin{align}
\hat{\mathbf{s}} = \left( {\begin{array}{c} \partial_s x \\ \partial_s y \end{array} } \right)
= \left( {\begin{array}{c} \sqrt{1-\partial_s y^2} \\ \partial_s y \end{array} } \right)
&= \left( {\begin{array}{c} 1-\frac{1}{2}\partial_s y^2 \\ \partial_s y \end{array} } \right) + O(|y|^4). \label{shat}\\
\hat{\mathbf{n}} &= \left({\begin{array}{c} -\partial_s y \\ 1-\frac{1}{2}\partial_s y^2 \end{array} } \right) + O(|y|^4).
\label{nhat}
\end{align}
\nn Using (\ref{Velnormdexp}) -- (\ref{nhat}),
\begin{align}
\widehat{\partial_t{\mathbf{X}}}\cdot \hat{\mathbf{s}} &=
1 - \frac{1}{2}\left(\partial_s y -\frac{ \partial_t y}{U}\right)^2 + O(|y|^4), \label{sv}\\
\widehat{\partial_t{\mathbf{X}}}\cdot \hat{\mathbf{n}} &= \left(\frac{\partial_t y}{U}-\partial_s y\right)
\left(1 - \frac{1}{2}\frac{\partial_t y^2}{U^2}\right) + \frac{\partial_t y}{U}
\left(-\frac{h}{U} - \frac{1}{2}\partial_s y^2\right) + O(|y|^5). \label{nv}
\end{align}

Using (\ref{sv}) and (\ref{nv}) we can write the force and torque balance equations
(\ref{fxb})--(\ref{torqueb}) and
the cost of locomotion $\eta$ in terms of $U$, $h$, and derivatives of $y$.

\subsection{Traveling-wave optimum \label{sec:trav}}

We first expand $\eta$ and use it to argue that the optimal shape dynamics is approximately
a traveling wave. This allows us to neglect certain terms, which simplifies the formulae for
the rest of the analysis. If the power expended at time $t$ is $P(t)$, the cost of locomotion is
\begin{align}
\eta &= \int_0^1 P(t) dt {\Bigg \slash} \int_0^1 U(t) dt. \\
&= \int_0^1 U(t) \int_0^1 \left[\left( \widehat{\partial_t{\mathbf{X}}}\cdot \hat{\mathbf{s}} \right)^2 +
\mu_t \left( \widehat{\partial_t{\mathbf{X}}}\cdot \hat{\mathbf{n}} \right)^2 \right]
\frac{\| \partial_t\mathbf{X} \|}{U(t)}\,ds\,dt {\Bigg \slash} \int_0^1 U(t)\,dt. \label{etaint}
\end{align}
\nn In (\ref{etaint}) we have assumed that the tangential motion of the snake is entirely forward,
with no portion moving backward, so the coefficient
in front of the tangential term is unity (i.e. $\mu_f$, rather than a term involving
both $\mu_f$ and $\mu_b$). This assumption is not required but it simplifies the equations from
this point onward and will be seen to be correct shortly.
The numerator of (\ref{etaint}) can be separated into a part involving tangential
velocity and a part involving normal velocity with a factor of $\mu_t$. Using
(\ref{Velnormexp}), (\ref{sv}), and (\ref{nv}), we can
list the terms with the lowest powers of $y$ in each part:
\begin{align}
\left( \widehat{\partial_t{\mathbf{X}}}\cdot \hat{\mathbf{s}} \right)^2
\frac{\| \partial_t\mathbf{X} \|}{U(t)} &= 1 + O(|y|^2) \label{sterm}\\
\mu_t \left( \widehat{\partial_t{\mathbf{X}}}\cdot \hat{\mathbf{n}} \right)^2
\frac{\| \partial_t\mathbf{X} \|}{U(t)} &=
\mu_t \left(\partial_s y -\frac{ \partial_t y}{U}\right)^2 + O(\mu_t |y|^4).\label{nterm}
\end{align}
\nn The optimal shape dynamics is a $y(s,t)$ that minimizes $\eta$
subject to the constraints of force and torque balance. The leading-order term
in (\ref{sterm}) is 1, which is independent of the shape dynamics. At
the next (quadratic) order in $y$, terms appear in both (\ref{sterm})
and (\ref{nterm}); only that in (\ref{nterm}) is given explicitly.
It is multiplied by $\mu_t$, so it appears to be
dominant in the large $\mu_t$ limit.
Thus, to a first approximation, minimizing $\eta$ is achieved by setting
\bqe
\left(\partial_s y -\frac{ \partial_t y}{U}\right) = 0,
\eqe
\nn so a first guess for the solution is a traveling wave, i.e. a function
of $s + Ut$ with $U$ constant. However, such a function does not satisfy the $x$-component of
the force balance equation (\ref{fxb}), which is:
\bqe
\int \left[\left( \widehat{\partial_t{\mathbf{X}}}\cdot \hat{\mathbf{s}} \right) s_x +
\mu_t \left( \widehat{\partial_t{\mathbf{X}}}\cdot \hat{\mathbf{n}} \right) n_x \right]\, ds = 0.
\label{fx}
\eqe
\nn Inserting (\ref{sv}) and (\ref{nv}) and retaining only the lowest powers
in $y$ from each expression,
\bqe
\int_0^1  1 + \mu_t \left(\partial_s y -\frac{ \partial_t y}{U}\right)\partial_s y\;ds = 0.
\label{fxexp}
\eqe
\nn which cannot be satistied by a function of $s + Ut$. Physically,
the first term (unity) in (\ref{fxexp}) represents the leading-order
drag force on the body due to forward friction. For a function of $s + Ut$,
in which the deflection wave speed equals the forward speed, the snake body
moves purely tangentially (up to this order of expansion in $y$)
along a fixed path on the ground, with no transverse
forces to balance the drag from tangential frictional forces. Also, when a function
of $s + Ut$ is inserted into (\ref{fxexp}) all factors of $U$ cancel out and
$U$ is left undetermined. This is related to a more fundamental
reason why posing the shape dynamics as a function of $s + Ut$ is invalid:
the problem of solving for the snake dynamics given in section \ref{sec:Model}
is not then well-posed. In the well-posed problem we provide the snake
curvature and curvature velocity versus time as inputs, and solve the force
and torque balance equations to obtain the average translational and rotational
velocities as outputs. Thus $U(t)$ is one of the three outputs (or unknowns)
that is needed to solve the three equations.

Instead we pose the shape dynamics as
\bqe
y(s,t) = g(s+ U_w t), \label{shapewave}
\eqe
which is a traveling wave with a prescribed wave speed $U_w$, different from
$U$ in general. Since $y$ is
periodic in $t$ with period 1, $g$ is periodic with a period of $U_w$. Inserting
(\ref{shapewave}) into (\ref{fxexp}), we obtain an equation for
$U$ in terms of $U_w$ and $g$:
\bqe
1 + \mu_t \left(1-\frac{U_w}{U}\right)\int_0^1 g'(s+U_w t)^2 ds = 0. \label{fxexpsol1}
\eqe
\nn In (\ref{fxexpsol1}), as $\mu_t \to \infty$, $(1-U_w/U) \to 0^-$. In the
solution given by (\ref{shapewave}) and (\ref{fxexpsol1}),
the shape wave moves backwards along the snake at speed $U_w$,
which propels the snake forward at a speed $U$, with $U$ slightly less than
$U_w$. Therefore the snake slips transversely to itself, in the backward
direction, which provides a forward thrust to balance the backward
drag due to forward tangential friction. We refer to $(1-U_w/U)$ in
(\ref{fxexpsol1}), a measure of the amount
of backward slipping of the snake, as the ``slip.'' 
In figure \ref{fig:HighMutPaperFig1}a,
the snake slips transversely to itself, and backward (leftward), even as
it moves rightward. Here the slip is fairly small because $\mu_t$
is 30, fairly large.

We now solve (\ref{fxexpsol1}) for $(1-U_w/U)$ and insert the result into
(\ref{etaint}) to find $g$ and $U_w$ which
minimize $\eta$. Using the lowest order expansions (\ref{sterm})
and (\ref{nterm}) with (\ref{fxexpsol1}) we obtain:

\begin{align}
\eta &= \int_0^1 P(t) dt {\Bigg \slash} \int_0^1 U(t) dt. \\
&= \int_0^1 U(t) \int_0^1
1 + \mu_t \left(\partial_s y -\frac{ \partial_t y}{U}\right)^2 + O(|y|^2, \mu_t |y|^4) ds dt
{\Bigg \slash} \int_0^1 U(t)\,dt. \\
&= 1 {\Bigg \slash} \int_0^1 \cfrac{1}{1+\cfrac{1}{\mu_t\langle g'(s+U_w t)^2 \rangle
 + O(|g|^2, \mu_t |g|^4)}} \,dt. \label{etaint1}
\end{align}
 \nn where
\bqe
\langle g'(s+U_w t)^2 \rangle \equiv \int_0^1 g'(s+U_w t)^2 ds.
\eqe
\nn If we minimize (\ref{etaint1}) over possible $g$
for a given $\mu_t$, using only the terms
in the expansion that are given explicitly, we find that $\eta$ tends to a minimum of 1
as the amplitude of $g$ diverges. However, in this limit our
small-$y$ power series expansions are no longer valid.
We therefore add the next-order terms in our expansions
of $\eta$ and the $F_x$ equation and search again for an
$\eta$-minimizer.

\subsection{Optimal wave amplitude \label{sec:amp}}

We now expand to higher order the components of $\eta$, (\ref{sterm})
and (\ref{nterm}),
\begin{align}
\left( \widehat{\partial_t{\mathbf{X}}}\cdot \hat{\mathbf{s}} \right)^2
\frac{\| \partial_t\mathbf{X} \|}{U(t)} &= 1
-\left(\partial_s y -\frac{ \partial_t y}{U}\right)^2
+ \frac{h}{U} + \frac{1}{2}\frac{\partial_t y^2}{U^2} + O(|y|^4) \label{sterm1}\\
\mu_t \left( \widehat{\partial_t{\mathbf{X}}}\cdot \hat{\mathbf{n}} \right)^2
\frac{\| \partial_t\mathbf{X} \|}{U(t)} &=
\mu_t \left[ \left(\frac{\partial_t y}{U} - \partial_s y\right) +
\frac{\partial_t y}{U}
\left(-\frac{h}{U} - \frac{1}{2}\partial_s y^2\right)
\right]^2 \left(1 + O(|y|^2)\right),\label{nterm1}
\end{align}
\nn and equation (\ref{fxb}),
\bqe
\int  1 + \mu_t \left(\partial_s y -\frac{ \partial_t y}{U} + \frac{\partial_t y}{U}
\left(\frac{h}{U} + \frac{1}{2}\partial_s y^2\right)\right)\partial_s y\;ds = 0.
\label{fxexpalt}
\eqe
\nn We again assume the traveling wave form of $y$ (\ref{shapewave}), and
use this to evaluate $h$ given by (\ref{h}) in terms of $g$:
\bqe
h(s,t) = -\frac{U_w}{2}g'(s+U_w t)^2 + \frac{U_w}{2}\langle g'(s+U_w t)^2 \rangle. \label{hg}
\eqe
\nn Inserting the traveling-wave forms of $y$ (\ref{shapewave}) and $h$
(\ref{hg}) into (\ref{fxexpalt}) we obtain
\bqe
1+\mu_t \left[\left(1-\frac{U_w}{U}\right)\langle g'(s+U_w t)^2 \rangle
+ \frac{1}{2}\langle g'(s+U_w t)^2 \rangle^2 \right] = 0. \label{fxexpaltg}
\eqe
\nn To obtain (\ref{fxexpaltg}) we've used the fact that the slip
$(1-U_w/U) \to 0$ as $\mu_t \to \infty$, which will be verified
subsequently. One can also proceed without this assumption, at the expense
of a lengthier version of (\ref{fxexpaltg}). We insert the traveling-wave
forms of $y$ and $h$ into (\ref{sterm1})
and (\ref{nterm1}) to obtain $\eta$, again with the assumption
that $(1-U_w/U) \to 0$ as $\mu_t \to \infty$ to simplify the result:
\bqe
\eta = 1 + \int_0^1 \frac{1}{2}\langle g'(s+U_w t)^2 \rangle +
\mu_t \left(1-\frac{U_w}{U}+ \frac{1}{2} \langle g'(s+U_w t)^2 \rangle \right)^2
\langle g'(s+U_w t)^2 \rangle\; dt \label{eta2g}
\eqe
\nn Solving for the slip $(1-U_w/U)$ from (\ref{fxexpaltg}) and inserting into
(\ref{eta2g}), we obtain an improved version of (\ref{etaint1}):
\bqe
\eta = 1 {\Bigg \slash} \int_0^1 \cfrac{1}{1+ \frac{1}{2}\langle g'(s+U_w t)^2 \rangle + \cfrac{1}{\mu_t\langle g'(s+U_w t)^2 \rangle}
 + O(|g|^4, \mu_t |g|^8)} \,dt. \label{etaint2}
\eqe
(\ref{etaint2}) is a more accurate version of (\ref{etaint1}), with an additional term
in the denominator of the integrand which penalizes large amplitudes for
$g$. If we approximate $\langle g'(s+U_w t)^2 \rangle$ as constant in time,
we obtain
\bqe
\eta = 1+ \frac{1}{2}\langle g'^2 \rangle +
\cfrac{1}{\mu_t\langle g'^2 \rangle}
 + O(|g|^4, \mu_t |g|^8). \label{etaint3}
\eqe
\nn which is minimized for
\bqe
\langle g'^2 \rangle^{1/2} = 2^{1/4} \mu_t^{-1/4}. \label{gopt}
\eqe
\nn Thus the RMS amplitude of the optimal $g$ should scale as $\mu_t^{-1/4}$.
(\ref{gopt}) represents a balance between competing effects. At smaller
amplitudes, the forward component of normal friction is not large enough
to balance the forward component of tangential friction, so the snake slips
more, which reduces its forward motion and does more work in the normal
direction. At larger amplitudes,
as already noted, the snake's path is more curved, which requires more
work done against tangential friction for a given forward distance traveled.
(\ref{gopt}) is only a criterion for the amplitude of the motion.
To obtain information about the shape of the optimal $g$,
we now consider the balances of the $y$-component of the force and the torque.

\subsection{Small-wavelength shapes \label{sec:small}}

So far we have searched for the optimal shape dynamics in terms of $y(s,t)$,
but in fact it is the curvature $\kappa(s,t)$ which is prescribed.
We obtain $y(s,t)$ from the curvature by integrating twice in $s$:
\begin{align}
y(s,t) &= y(0,t) + \int_0^s \sin \theta(s',t) ds' \\
&= y(0,t) + \int_0^s \theta(s',t) ds' + O(y^3) \\
&= y(0,t) + \int_0^s \left[ \theta(0,t) +
\int_0^{s'} \kappa(s'',t) ds'' \right] ds' + O(y^3).\\
&= y(0,t) + \theta(0,t) s + \int_0^s
\int_0^{s'} \kappa(s'',t) ds'' ds' + O(y^3).\\
&\equiv Y(t) + s R(t) + k(s,t)+ O(y^3). \label{newy}
\end{align}
\nn where $Y$ and $R$ are defined for notational convenience.
Prescribing the curvature is equivalent to prescribing $k(s,t)$. We set
\bqe
k(s,t) = g(s + U_w t) \label{k}
\eqe
\nn so we have the same form for $y$ as before, with an additional
translation $Y(t)$ and rotation $R(t)$. $Y$ and $R$ are determined by
the $y$-force and torque balance equations:
\begin{align}
\int_0^1 \left[\left( \widehat{\partial_t{\mathbf{X}}}\cdot \hat{\mathbf{s}} \right) s_y +
\mu_t \left( \widehat{\partial_t{\mathbf{X}}}\cdot \hat{\mathbf{n}} \right) n_y \right]\, ds = 0.
\label{fy} \\
\int_0^1 \left[\left( \widehat{\partial_t{\mathbf{X}}}\cdot \hat{\mathbf{s}} \right)
(x s_y - y s_x) +
\mu_t \left( \widehat{\partial_t{\mathbf{X}}}\cdot \hat{\mathbf{n}} \right) (x n_y - y n_x) \right]\, ds = 0.
\label{torque}
\end{align}
\nn We expand these to leading order in $y$ and obtain
\begin{align}
\int_0^1 \partial_s y -\frac{ \partial_t y}{U} + \frac{\partial_t y}{U}
\left(\frac{h}{U} + \frac{1}{2}\partial_s y^2\right)\,ds = 0.
\label{fy1} \\
\int_0^1 s\left(\partial_s y -\frac{ \partial_t y}{U} + \frac{\partial_t y}{U}
\left(\frac{h}{U} + \frac{1}{2}\partial_s y^2\right)\right)\,ds = 0.
\label{torque1}
\end{align}
We insert (\ref{newy}) with (\ref{k}) into the three equations (\ref{fxexpalt}), (\ref{fy1}),
and (\ref{torque1}) to solve for the three unknowns, $U$, $Y$, and $R$ in terms of $g$ and
$U_w$. We obtain:
\begin{align}
\int_0^1 1 + \mu_t\left[\left(-\frac{U_w}{U}+1+\frac{1}{2}\langle g'^2\rangle\right)g'^2
+\left(-\frac{Y'}{U} - \frac{R's}{U} + R\right)g' \right] ds = 0. \label{fx2}\\
\int_0^1 \left(-\frac{U_w}{U}+1+\frac{1}{2}\langle g'^2\rangle\right)g'
-\frac{Y'}{U} - \frac{R's}{U} + R\, ds = 0. \label{fy2}\\
\int_0^1 s \left[\left(-\frac{U_w}{U}+1+\frac{1}{2}\langle g'^2\rangle\right)g'
-\frac{Y'}{U} - \frac{R's}{U} + R\right]\, ds = 0.\label{torque2}
\end{align}
\nn We solve (\ref{fy2}) and (\ref{torque2}) for $Y$ and $R$ in terms of $U$:
\begin{align}
\frac{Y'}{U}-R &= 4B - 6A \label{Y1} \\
\frac{R'}{U} &= 12A - 6B \\
A &\equiv \left(-\frac{U_w}{U}+1+\frac{1}{2}\langle g'^2\rangle\right)\langle sg'\rangle \\
B &\equiv \left(-\frac{U_w}{U}+1+\frac{1}{2}\langle g'^2\rangle\right)\langle g'\rangle \label{B}
\end{align}
\nn We then solve (\ref{fx2}) for $U$ in terms of $g$:
\bqe
\frac{U_w}{U}= 1+\frac{1}{2}\langle g'^2\rangle + \frac{1}{\mu_t\left(
\langle g'^2\rangle - \langle g'\rangle^2 -
3\left(\langle g'\rangle - 2\langle s g'\rangle \right)^2\right)}. \label{U2}
\eqe
\nn We now form $\eta$, (\ref{etaint}) with terms
given by (\ref{sterm1}) and (\ref{nterm1}), using the updated from of $y$
((\ref{newy}) with (\ref{k})) and (\ref{U2}). We obtain a more
correct version of
(\ref{etaint2}):
\bqe
\eta = 1 {\Bigg \slash} \int_0^1 \cfrac{1}{1+ \frac{1}{2}\langle g'^2 \rangle +
\cfrac{1}{\mu_t\left(
\langle g'^2\rangle - \langle g'\rangle^2 -
3\left(\langle g'\rangle - 2\langle s g'\rangle \right)^2\right)}
 + O(|g|^4, \mu_t |g|^8)} \,dt. \label{etaint4}
\eqe
\nn We can find $\eta$-minimizing $g$ in a few steps.
First, let $\{L_k\}$ be the family of orthonormal polynomials
with unit weight on $[0,1]$ (essentially the Legendre polynomials):
\bqe
\int_0^1 L_i L_j \, ds = \delta_{ij} \quad ;\quad L_0 \equiv 1 \;,\; L_1 = \sqrt{12}(s - 1/2)
\;,\ \ldots.
\eqe
\nn At a fixed time $t$ we expand $g'$ in the basis of the $L_k$:
\bqe
g'(s+U_w t) = \sum_{k=0}^\infty c_k(t) L_k(s).
\eqe
\nn Then we have:
\begin{align}
\langle g'^2\rangle &= \sum_{k=0}^\infty c_k(t)^2, \label{ident1}\\
\langle g'\rangle^2 &= c_0(t)^2, \\
3\left(\langle g'\rangle - 2\langle s g'\rangle \right)^2 &= \langle L_1 g'\rangle^2 = c_1(t)^2. \label{ident3}
\end{align}
\nn Inserting into (\ref{etaint4}), $\eta$ becomes
\bqe
\eta = 1 {\Bigg \slash} \int_0^1 \cfrac{1}{1+ \frac{1}{2}
\left(c_0(t)^2 + c_1(t)^2 + \sum_{k=2}^\infty c_k(t)^2\right) +
\cfrac{1}{\mu_t\left(
\sum_{k=2}^\infty c_k(t)^2 \right)}
 + O(|g|^4, \mu_t |g|^8)} \,dt. \label{etaint5}
\eqe
\nn The denominator of the integrand in (\ref{etaint5}) is
minimized when
\bqe
c_0(t) = 0 \; ; \; c_1(t) = 0 \; ; \; \sum_{k=2}^\infty c_k(t)^2 = \sqrt{2}\mu_t^{-1/2}. \label{min}
\eqe
\nn The function $g(s+U_w t)$ that minimizes $\eta$ is that for which
(\ref{min}) holds for all $t$. Then the integral in (\ref{etaint5}) is
maximized, so $\eta$ is minimized. Recall that $g$ is a periodic function
with period $U_w$. The relations (\ref{min}) hold for {\it any} such
$g$ in the limit that $U_w \to 0$, as long as $g$ is normalized
appropriately. Define
\bqe
A \equiv \left(\frac{1}{U_w}\int_0^{U_w} g'(x)^2 dx\right)^{1/2}. \label{A}
\eqe
\nn Then
\begin{align}
\int_0^1 g'(s + U_w t)^2 ds &= A^2 + O(U_w) \label{ms}\\
\int_0^1 g'(s + U_w t) ds &= O(U_w) \label{avg}\\
\int_0^1 s g'(s + U_w t) ds &= O(U_w). \label{moment}
\end{align}
\nn (\ref{ms}) -- (\ref{moment}) are fairly straightforward to show, and
we show them in Appendix \ref{sec:smallint} for completeness. If 
(\ref{ms}) -- (\ref{moment}) hold, then in the limit that
$U_w \to 0$, (\ref{min}) holds with $A = 2^{1/4}\mu_t^{-1/4}$, by 
(\ref{ident1}) -- (\ref{ident3}). A physical interpretation of
the small-wavelength limit is that the net vertical force and
torque on the snake due to the traveling wave alone ($g$) become zero in
this limit, so no additional heaving motion ($Y$) or rotation
($R$) are needed, and thus the additional work associated
with these motions is avoided. Similar ``recoil" conditions were
proposed as kinematic constraints in the context of fish 
swimming \cite{lighthill1975mathematica}. 
We also note two other scaling laws. For a given $g \sim \mu_t^{-1/4}$, by
(\ref{U2}) the slip $(1-U_w/U) \sim \mu_t^{-1/2}$
and by (\ref{Y1})--(\ref{B}), $Y, R \sim \mu_t^{-3/4}$.

\subsection{Optimal cost of locomotion and comparison with numerics \label{sec:comp}}

To obtain an $\eta$-minimizing function $g^*$, let $g$ be any periodic function
with period $U_w$. Define
\bqe
g^*(s+U_w t) \equiv \frac{2^{1/4}\mu_t^{-1/4}}{A} g(s+U_w t)
\eqe
\nn where $A$ is given in (\ref{A}). Then in the limit $U_w \to 0$,
\bqe
\eta(g^*) \to 1 + \sqrt{2}\mu_t^{-1/2} + O(\mu_t^{-1}), \label{etaopt1}
\eqe
\nn by the estimate in (\ref{etaint4}). Thus the optimal traveling wave motions
become more efficient as $\mu_t \to \infty$, which agrees with
the numerical solutions in \cite{AlbenSnake2013I}.
The numerically-determined optima in figure \ref{fig:HighMutPaperFig1} do not
have small wavelengths, due to the finite number of modes used. This
is explained further in the Supplementary Material of \cite{AlbenSnake2013I}.

\begin{figure} [h]
           \begin{center}
           \begin{tabular}{c}
               \includegraphics[width=6in]{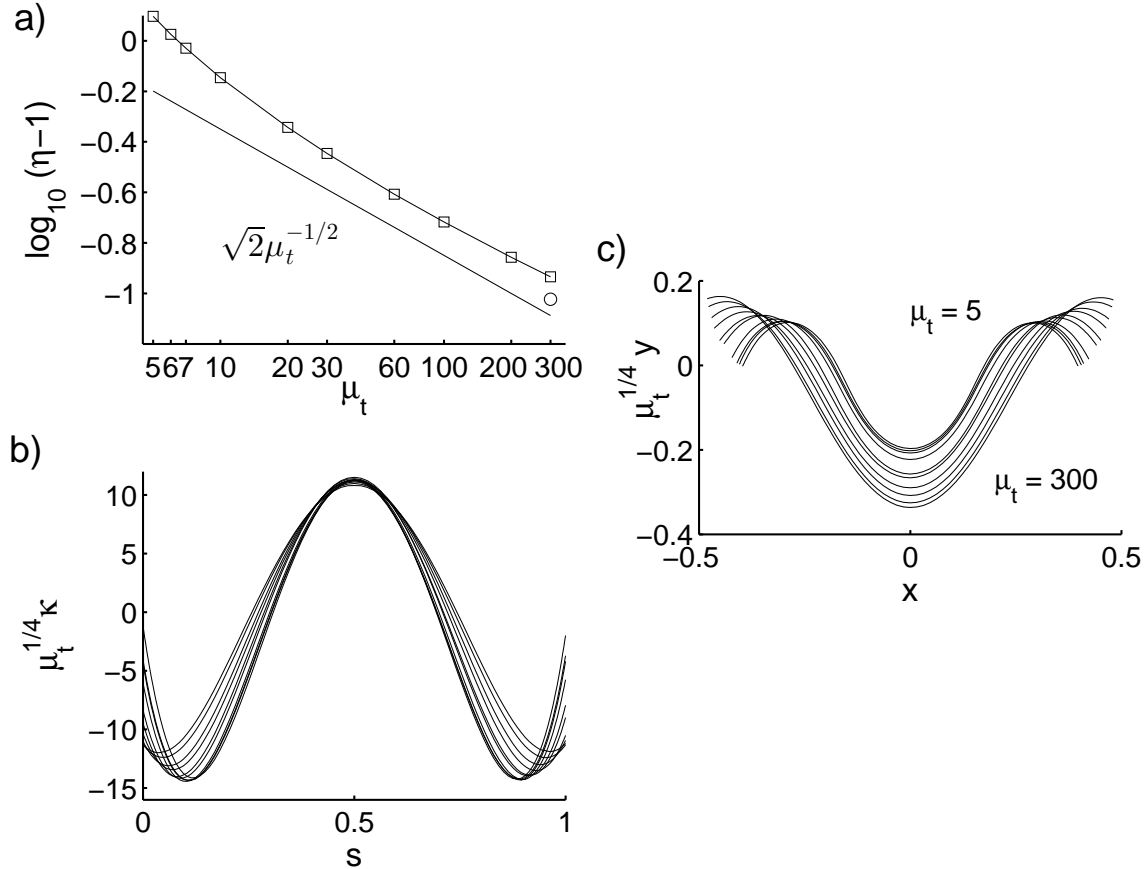} \\
           \vspace{-.25in}
           \end{tabular}
          \caption{\footnotesize The $\mu_t$-scalings of
the numerical traveling-wave optima
from figure \ref{fig:HighMutPaperFig1} and \cite{AlbenSnake2013I}.
(a) The cost of locomotion $\eta$ minus
unity, on a log scale, for the numerical optima (squares) of
\cite{AlbenSnake2013I} together with $\eta-1$ for the analytical
optimum (solid line). (b) The curvature from figure \ref{fig:HighMutPaperFig1}b
rescaled by $\mu_t^{1/4}$. (c) The deflection from
figure \ref{fig:HighMutPaperFig1}c rescaled by $\mu_t^{1/4}$.
 \label{fig:RescaledHighMutFig}}
           \end{center}
         \vspace{-.10in}
        \end{figure}

We compare the numerically-found optima to the theoretical
results in figure \ref{fig:RescaledHighMutFig}. Here we take
the plots of figure \ref{fig:HighMutPaperFig1} and transform them according
to the theory. In figure
\ref{fig:RescaledHighMutFig}a
we plot $\eta-1$ versus $\mu_t$ for the numerical solutions (squares)
from \cite{AlbenSnake2013I}, along
with the small-$U_w$ theoretical result  (solid line), which is given
by (\ref{etaopt1}).
The numerics
follow the same scaling as the theory, but with a consistent upward shift.
Some degree of upward shift is expected from the finite-mode truncation
in the numerical computation, which
makes the numerical result underperform the analytical result, as
explained in \cite{AlbenSnake2013I}. The circle shows the efficiency
for a curvature function with a higher wavelength than those which
could be represented in the numerical optimization, and its distance from
the solid line is $O(\mu_t^{-1})$, of the order of the next term in the
asymptotic expansion of $\eta - 1$.

Figure \ref{fig:RescaledHighMutFig}b shows the curvature, rescaled by $\mu_t^{1/4}$ to
give a collapse according to (\ref{gopt}). The data collapse well compared
to the unscaled data in figure \ref{fig:HighMutPaperFig1}b. 
Figure \ref{fig:RescaledHighMutFig}c
shows the body shapes from figure \ref{fig:HighMutPaperFig1}c 
with the vertical coordinate rescaled, and the
shapes plotted with all centers of mass located at the origin. 
We again find a good collapse,
consistent with the curvature collapse.

\section{Conclusion \label{sec:Concl}}

We have studied the optimization of planar
snake motions for efficiency, using a fairly simple model
for the motion of snakes using friction. When
the coefficient of transverse friction is
large, our analysis
shows that a traveling-wave motion of small amplitude
is optimal. The amplitude tends to zero as the
transverse friction coefficient tends to infinity, scaling
as the transverse friction coefficient to the
-1/4 power, and the efficiency tends to unity in this limit. 
The corresponding power is that for a straight snake towed forward.

In \cite{AlbenSnake2013I} we were able to compute
many optimal motions at moderate and small $\mu_t$ also.
Those found at small $\mu_t$ also corresponded to traveling
waves (direct now, not retrograde), and at zero $\mu_t$,
a simple triangular-wave solution was found with
optimal efficiency. It remains to determine optimal motions
with small but nonzero $\mu_t$ analytically.
At moderate $\mu_t$, a region of standing-wave or ratcheting
motions was found. It may be possible to compute some of
these motions analytically using a set of assumptions specialized
to the moderate-$\mu_t$ regime.

\begin{acknowledgments}
We would like to acknowledge helpful discussions on snake
physiology and mechanics with David Hu and Hamidreza Marvi,
helpful discussions with Fangxu Jing during our previous
study of two- and three-link snakes, and
the support of NSF-DMS
Mathematical Biology Grant 1022619 and a Sloan Research Fellowship.
\end{acknowledgments}

\appendix

\section{Small-wavelength integrals \label{sec:smallint}}
To show (\ref{ms}) -- (\ref{moment}), we first note that
since $g$ is periodic, $g'$ is periodic with zero average over a period.
To show (\ref{ms}), we decompose the integral in (\ref{ms}) into two intervals:
\bqe
\int_0^1 g'(s + U_w t)^2 ds = \int_0^{U_w\lfloor{1/U_w}\rfloor} g'(s + U_w t)^2 ds + \int_{U_w\lfloor{1/U_w}\rfloor}^1 g'(s + U_w t)^2 ds. \label{msarg}
\eqe
\nn The first integral on the right of (\ref{msarg}) is over an integral number
of periods of $g'^2$, and is thus $\lfloor{1/U_w}\rfloor U_w A^2 = A^2 + O(U_w)$.
The second integral is over an interval of length $< U_w$, and is thus $O(U_w)$,
which shows (\ref{ms}).
To show (\ref{avg}), we use (\ref{msarg}) again but with $g'$ in place of $g'^2$:
\bqe
\int_0^1 g'(s + U_w t) ds = \int_0^{U_w\lfloor{1/U_w}\rfloor} g'(s + U_w t) ds + \int_{U_w\lfloor{1/U_w}\rfloor}^1 g'(s + U_w t) ds. \label{avgarg}
\eqe
The first integral on the right of (\ref{avgarg}) is over an integral number
of periods of $g'$, and is thus zero. The second integral is over an interval of
length $< U_w$, and is thus $O(U_w)$. To show (\ref{moment}), we decompose the integral
into three parts:
\begin{align}
\int_0^1 s & g'(s + U_w t) ds = \int_0^{U_w\lfloor{1/U_w}\rfloor}
{U_w\lfloor{s/U_w}\rfloor} g'(s + U_w t) ds \no \\
&+ \int_0^{U_w\lfloor{1/U_w}\rfloor} (s - {U_w\lfloor{s/U_w}\rfloor})g'(s + U_w t) ds
+  \int_{U_w\lfloor{1/U_w}\rfloor}^1 s g'(s + U_w t) ds. \label{momentarg}
\end{align}
The first integral on the right of (\ref{momentarg}) substitutes a constant
approximation to $s$ on each subinterval of period length. It is identically zero
since $g'$ has mean zero.
The second integral is the error in the constant approximation to $s$, $O(U_w)$
on each subinterval, of which there are $O(1/U_w)$, each of length $U_w$.
Thus the second integral is $O(U_w)$. The third integral is also $O(U_w)$, as
explained for (\ref{msarg}) and (\ref{avgarg}), so (\ref{moment}) holds.

\bibliographystyle{unsrt}
\bibliography{snake}

\end{document}